\documentstyle[aps,epsf]{revtex}
\begin{document}
\draft
\title{Spin dynamics in nonlinear optical spectroscopy of
Fermi sea systems}
\author{T. V. Shahbazyan}
\address{Department of Physics and Astronomy,  
Vanderbilt University, Nashville, TN 37235}
\author{I. E. Perakis}
\address{Department of Physics, University of Crete,
P.O. Box 2208, 710 03, Heraklion, Crete, Greece}
\author{M. E. Raikh}
\address{Department of Physics, University of Utah, Salt Lake City,
UT 84112}
\maketitle

\begin{abstract}    
We discuss the role of many-body spin correlations in nonlinear optical
response of a Fermi sea system with a deep impurity level. 
Due to the Hubbard repulsion between electrons at the impurity, the
optical transitions between the impurity level and the Fermi sea states
lead to an optically-induced Kondo effect.
In particular, the third-order nonlinear optical susceptibility
logarithmiclly diverges at the absorption threshold.
The shape of the pump-probe spectrum is governed by the
light-induced  Kondo temperature, which can be tuned by
varying the intensity and  frequency of the pump optical field. 
In the Kondo limit, corresponding to off-resonant pump excitation, 
the nonlinear absorption spectrum exhibits  a narrow peak below the
linear absorption onset. 
\end{abstract}
%\pacs{PACS numbers:  78.47.+p 42.65 -k, 72.15.Qm}

%\vspace{0.5in}

% Keywords: spin dynamics, ultrafast nonlinear optics, Kondo effect
%\narrowtext

%\clearpage

\section{introduction}

The role of many-body correlations in the nonlinear optical
spectroscopy of Fermi sea systems has attracted much interest
during the last decade\cite{chemla99,shah96,perakis00}.
Due to the development of high quality ultrashort laser pulses,
it has become possible to probe the elementary excitations of an
interacting system on time scales shorter than the 
dephasing and relaxation times. 
In undoped semiconductors, it has been established that
exciton-exciton interactions  play a dominant role in the coherent 
regime\cite{chemla99}.
Very recently, there has been
a growing interest in studying the {\em coherent} dynamics 
of  Fermi sea (FS) systems at low
temperatures\cite{bre95,portengen,awschalom,hall,dodge99,FES}.  
% In strongly-correlated systems,
% such as quantum Hall or magnetic systems, the many-body correlations
% dominate the time evolution of the nonlinear optical 
% polarization\cite{hall,dodge99}.
% In modulation-doped quantum wells,
% it was shown that, for below-resonance 
% photoexcitation,  the dynamics of the Fermi edge singularity
% is determined by the characteristic Coulomb
% energy of the low-lying FS excitations\cite{FES,bre95}. 

In this paper, we discuss a new many-body effect in the nonlinear
optical response of a  FS system with a deep impurity level. 
Specifically, we will focus on nonlinear absorption due to optical
transitions between a localized impurity level  
and the continuum of FS states.
In the linear  absorption, two prominent many-body effects has long
been known in such systems\cite{mahan-book}. First is the Mahan 
singularity due to the attractive
interaction between the FS and the localized hole. 
Second is the Anderson orthogonality
catastrophe due to the readjustment 
of the FS density profile during the optical 
transition.

The nonlinear absorption comes from {\em multiple} transitions between
impurity level and the Fermi sea. In particular, a number of different 
intermediate processes contribute to the third-order 
optical susceptibility $\chi^{(3)}$\cite{mukamel-book}. 
What is crucial for us here is that, in the system under study,
some of the intermediate states involve the doubly-occupied impurity
level. For example, the optical field can first cause a transition 
of a FS electron  to the singly-occupied impurity level, 
which thus becomes doubly-occupied, and then excite both electrons
from the impurity level to the conduction band. 
This is illustrated in Fig.\ 1(a). Important is that, while
on the impurity, the two electrons experience a Hubbard repulsion. 
In fact, such a repulsion gives leads to a logarithmic divergence in
$\chi^{(3)}$. The origin of such an anomaly is intimately related to the
Kondo effect\cite{shahbazyan00}. 

% Furthermore, for for off-resonant pump,
% the nonlinear absorption spectrum exhibits a narrow peak below the
% linear absorption onset. 

To be specific, we restrict ourselves to 
pump-probe spectroscopy, where 
a strong pump and a weak probe optical field
are applied to the system and  the optical 
polarization along the probe direction is measured. 
We only consider near-threshold absorption at zero temperature
and assume that the pump frequency is tuned below 
the onset of optical transitions from the impurity 
level so that dephasing processes due to electron-electron and
electron-phonon interactions are suppressed.

We identify two distinct regimes which are characterized by the
interplay between pump detuning and pump intensity. 
These regimes are somewhat analogous to the Kondo limit and
mixed-valence regime in the ``usual'' Kondo systems. The 
shapes of nonlinear absorption spectra in these regimes are
drastically different. The crossover
between the two regimes is governed by a new energy scale -- the
liht-induced Kondo temperature, which can be tuned by varying the
intensity and frequency of the pump.

The paper is organized as follows. In Section II, we derive the
Kondo-induced contribution to the third-order optical polarization
that determines the absorption spectrum. In Section III, we obtain
the absorption coefficient beyond $\chi^{(3)}$ using variational large
$N$ method. Section IV concludes the paper.

\begin{figure}
\begin{center}
\epsfxsize=4.0in
\epsffile{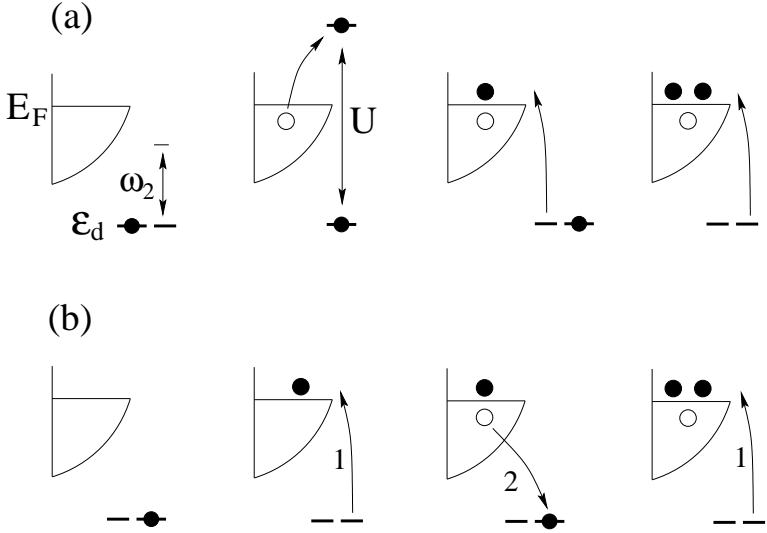}
\end{center}
%\vspace{0.5in}

\caption{
Intermediate processes contributing to $\chi^{(3)}$. 
(a) Intermediate state with doubly-occupied impurity: 
excitation of a FS electron to the second impurity state followed by
the excitation of the two impurity electrons to the FS. 
(b) Large $U$ limit: {\em Two} transition channels are possible from
states {\em below} the FS to the two {\em spin-degenerate} states of 
empty impurity, but only {\em one} channel from the 
{\em singly-occupied} impurity to states {\em above} the FS. 
}
\end{figure}

\section{Kondo anomaly in $\chi^{(3)}$}

We start with the Hamiltonian of the system:
$H_{tot}=H+H_{1}(t)+H_{2}(t)$, where 
\begin{equation}
\label{Ham}
H=\sum_{{\bf k}\sigma}\varepsilon_kc_{{\bf k}\sigma}^{\dag}c_{{\bf k}\sigma}
+\varepsilon_d\sum_{\sigma}d_{\sigma}^{\dag}d_{\sigma}
+\frac{U}{2}\sum_{\sigma\neq\sigma'}\hat{n}_{\sigma}\hat{n}_{\sigma'},
\end{equation}
is the Hamiltonian in the absence of optical fields; here 
$c_{{\bf k}\sigma}^{\dag}$ and $d_{\sigma}^{\dag}$ are conduction 
and localized electron creation operators, respectively,
($\hat{n}_{\sigma}=d_{\sigma}^{\dag}d_{\sigma}$);
$\varepsilon_k$ and $\varepsilon_d$ are the corresponding energies,
and $U$ is the Hubbard interaction (all energies are measured from the
Fermi level). 
The coupling to the optical fields is described by the Hamiltonian
\begin{equation}
\label{Hi}
H_i(t)=-M_i(t)\hat{T}^{\dag} + {\rm h.c.},
~~~
\hat{T}^{\dag}=
\sum_{{\bf k}\sigma}c_{{\bf k}\sigma}^{\dag}d_{\sigma},
~~~
M_i(t)=e^{i{\bf k}_i\cdot{\bf r}-i\omega_it}\mu{\cal E}_i(t),
\end{equation}
where $\hat{T}^{\dag}$ is the optical transition operator.
%$M_i(t)=e^{i{\bf k}_i\cdot{\bf r}-i\omega_it}\mu{\cal E}_i(t)$
Here ${\cal E}_i(t)$, ${\bf k}_i$, and $\omega_i$ are the 
pump/probe electric field amplitude, direction and 
central frequency, respectively, and $\mu$ is the dipole
matrix element ($i=1,2$ denotes the probe and pump, respectively).
The pump-probe polarization is obtained
by expanding the optical polarization, $\mu\,\langle\hat{T}\rangle$, to
the first order in $H_1$ and keeping the terms propagating in the
probe direction\cite{mukamel-book}:
\begin{eqnarray}\label{PP}
P(t)=
i\mu \int_{-\infty}^{t} dt' 
%&&
M_1(t') 
\biggl[\langle\Phi(t)|\hat{T}{\cal K}(t,t')
\hat{T}^{\dag} |\Phi(t')\rangle
%\nonumber\\&&
-\langle\Phi(t')|\hat{T}^{\dag} {\cal K}(t',t)
\hat{T}|\Phi(t)\rangle\biggr],
\end{eqnarray}
where ${\cal K}(t,t')$ is the evolution operator for the Hamiltonian
$H+H_2(t)$ and the state 
$|\Phi(t)\rangle$  satisfies  the Schr\"{o}dinger equation 
$i\partial_t |\Phi(t)\rangle=[H+H_2(t)]|\Phi(t)\rangle$. 

The third order polarization is obtained by expanding  
${\cal K}(t,t')$ and $|\Phi(t)\rangle$ 
up to the second order in $H_2$. Below we consider sufficiently large
values of $U$ so that, in the absence of optical fields, the ground
state of $H$, $|\Omega_0\rangle$,    
represents a {\em singly-occupied} impurity  and  full FS.
For large $U$, the doubly-occupied impurity states are energetically
unfavorable and can be excluded from the expansion of the
polarization (\ref{PP}) with respect to $H_2$. 
% %
% Correspondingly, the processes involving two successive
% absorptions of a pump photon do not contribute.
%
The third-order pump-probe polarization then
takes the form 
$P^{(3)}(t)=e^{i{\bf k}_1 \cdot{\bf r}-i\omega_1t}\tilde{P}^{(3)}$ with
\begin{eqnarray}\label{PP3}
\tilde{P}^{(3)}=i\mu^4
\int_{-\infty}^{t} dt' {\cal E}_1(t')e^{i\omega_1(t-t')}
\biggl[Q_1(t,t')+Q_1^{*}(t',t)+Q_2(t,t')+Q_3(t,t')\biggr],
\end{eqnarray}
where
\begin{eqnarray}\label{Q}
&&
Q_1(t,t')=
-\int_{-\infty}^{t'} dt_1\int_{-\infty}^{t_1} dt_2
f(t_1,t_2)F(t,t',t_1,t_2),
\nonumber\\&&
Q_2(t,t')=
-\int_{t'}^{t} dt_2\int_{t'}^{t_2} dt_1
f(t_1,t_2)
F(t,t_2,t_1,t'),
\nonumber\\&&
Q_3(t,t')=
-\int_{-\infty}^{t'} dt_1\int_{-\infty}^{t} dt_2
f(t_1,t_2)
F(t_1,t',t,t_2).
\end{eqnarray}
Here we denoted 
$f(t_1,t_2)={\cal E}_2(t_1){\cal E}_2(t_2)e^{i\omega_2(t_1-t_2)}$, 
and
\begin{eqnarray}\label{F}
F(t,t',t_1,t_2)=&&
\langle \Omega_0|\hat{T}e^{-iH(t-t')}\hat{T}^{\dag}e^{-iH(t'-t_1)}
\hat{T}e^{-iH(t_1-t_2)}\hat{T}^{\dag}|\Omega_0\rangle
\nonumber\\
=&&
\sum_{{\bf pq}{\bf k}'{\bf k}\lambda s \sigma' \sigma}
A_{{\bf pq}{\bf k}'{\bf k}}^{\lambda s \sigma' \sigma}
%\nonumber\\&&\times
e^{-i(\varepsilon_p-\varepsilon_d)(t-t')
-i(\varepsilon_k-\varepsilon_{k'})(t'-t_1)
-i(\varepsilon_k-\varepsilon_d)(t_1-t_2)},
\end{eqnarray}
\begin{eqnarray}
\label{relations1}
A_{{\bf pq}{\bf k}'{\bf k}}^{\lambda s \sigma' \sigma}=
\langle\Omega_0|d_{\lambda}^{\dag}c_{{\bf p}\lambda}
c_{{\bf q} s}^{\dag}d_s
d_{\sigma'}^{\dag}c_{{\bf k}'\sigma'}
c_{{\bf k}\sigma}^{\dag}d_{\sigma}|\Omega_0\rangle
=
%&&
\delta_{\lambda \sigma}\delta_{s\sigma'}
n_{\sigma}(1-n_p)
%\nonumber\\&&\times
[\delta_{{\bf pk}}\delta_{{\bf qk}'}n_q
+\delta_{\sigma \sigma'}\delta_{{\bf pq}}\delta_{{\bf kk}'}(1-n_k)],
\end{eqnarray}
with 
$n_{\sigma}=\langle \Omega_0|d_{\sigma}^{\dag}d_{\sigma}|\Omega_0\rangle$ 
and $n_k=\langle \Omega_0|c_{{\bf k}\sigma}^{\dag}c_{{\bf k}\sigma}
|\Omega_0\rangle$ (impurity occupation number is 
$n_d=\sum_{\sigma}n_\sigma=1$ here).
Eqs.\ (\ref{F})\ and\  (\ref{relations1})
%, being $U$-independent,
are valid in the large $U$ limit corresponding to a singly-occupied
impurity.
For monochromatic optical fields, ${\cal E}_i(t)={\cal E}_i$, the
time integrals can be explicitly evaluated,
\begin{eqnarray}\label{Q2}
%&&
Q_1(t,t')&&+Q_1^{*}(t,t')=
{\cal E}_2^2\sum_{\bf pq}e^{-i(\epsilon_p-\epsilon_d)(t-t')}(1-n_p)
\Biggl[
\frac{2Nn_q}{(\epsilon_p-\epsilon_q)(\epsilon_p-E_d)}
-\frac{i(t-t')(1-n_q)}{(\epsilon_q-E_d)}\Biggr],
\nonumber\\
%&&
Q_2(t,t')&&=
{\cal E}_2^2\sum_{\bf pq}e^{-i(\epsilon_p-\epsilon_d)(t-t')}(1-n_p)
\Biggl[
-\frac{i(t-t')Nn_q}{(\epsilon_q-E_d)}
+\frac{2(1-n_q)}
{(\epsilon_p-\epsilon_q)(\epsilon_p-E_d)}
+\frac{(1-n_q)e^{i(\epsilon_p-E_d)(t-t')}}
{(\epsilon_q-E_d)(\epsilon_p-E_d)}
\Biggr],
\nonumber\\
%&&
Q_3(t,t')&&
=-{\cal E}_2^2\sum_{\bf pq}(1-n_p)
\Biggl[\frac{e^{i(\epsilon_p-\epsilon_q-\omega_2)(t-t')}Nn_q+
e^{-i\omega_2(t-t')}(1-n_q)}
{(\epsilon_q-E_d)(\epsilon_p-E_d)}
\Biggr],
\end{eqnarray}
yielding $\tilde{P}^{(3)}=\tilde{P}^{(3)}_0+\tilde{P}^{(3)}_K$ 
with   
\begin{eqnarray}
\tilde{P}_0^{(3)}=&&
\mu^4{\cal E}_1{\cal E}_2^2\sum_{\bf pq}
\frac{(1-n_p)}{\varepsilon_p-\varepsilon_d-\omega_1}
\Biggl[
\frac{2}{(\varepsilon_p-\varepsilon_q)(\varepsilon_p-E_d)}
-\frac{1}{(\varepsilon_p-\varepsilon_d-\omega_1)(\varepsilon_q-E_d)}
\Biggr],\label{pol-free}
\\
\tilde{P}_{K}^{(3)}=&&
(N-1)\mu^4 {\cal E}_1{\cal E}_2^2\sum_{\bf pq}
\frac{(1-n_p)n_q}{\varepsilon_p-\varepsilon_d-\omega_1}
\Biggl[
\frac{2}{(\varepsilon_p-\varepsilon_q)(\varepsilon_p-E_d)}
-\frac{1}{(\varepsilon_p-\varepsilon_d-\omega_1)(\varepsilon_q-E_d)}
\Biggr],\label{pol-kondo}
\end{eqnarray}
where $N$ is the impurity level degeneracy.
Here we introduced the effective impurity level
$E_d=\varepsilon_d+\omega_2$. 
The first term, $\tilde{P}_0^{(3)}$, is the usual third-order
polarization for {\em spinless} ($N=1$)
electrons\cite{mukamel-book}.  The second
term, $\tilde{P}_{K}^{(3)}$, originates from the
suppression, due to the Hubbard repulsion $U$,
of the contributions from   
doubly-occupied impurity states. 
As indicated by the prefactor $(N-1)$,  
it comes from the additional intermediate states
that are absent in the spinless case  [see Fig 1(b)].

Consider the first term in Eq. (\ref{pol-kondo}).
The restriction of the sum over ${\bf q}$ to states {\em below} the
Fermi level results in a logarithmic divergence
in the absorption  coefficient, 
$\alpha\propto {\rm Im}\tilde{P}$, 
at the absorption threshold, $\omega_1=-\varepsilon_d$:
\begin{equation}
\label{alpha3}
{\rm Im}\tilde{P}_K^{(3)}=(N-1)p_0\theta(\omega_1+\varepsilon_d)
\frac{2\Delta}{\pi\delta\omega} 
\ln \biggl|\frac{D}{\omega_1+\varepsilon_d}\biggr|,
\end{equation}
where $p_0=\pi{\cal E}_1\mu^2g$, 
$\delta\omega=\omega_1-\omega_2$ is the pump-probe detuning, and
$\Delta=\pi g \mu^2 {\cal E}_2^2$
is the energy width characterizing the pump intensity;
$D$ and $g$ are the bandwidth and the density of states (per
spin) at the Fermi level, respectively.  
Recalling that the linear absorption is determined by 
${\rm Im}\tilde{P}^{(1)}=p_0 \theta(\omega_1+\varepsilon_d)$,
we  see that it differs from Eq.\ (\ref{alpha3}) by  a factor
$\frac{2\Delta}{\pi\delta\omega}
\ln \bigl|\frac{D}{\omega_1+\varepsilon_d}\bigr|$
(setting for simplicity $N=2$).
In other words,  ${\rm Im}\tilde{P}^{(1)}$ and 
${\rm Im}\tilde{P}^{(3)}_K$ become comparable when
\begin{equation}
\label{break}
\omega_1+\varepsilon_d\equiv\delta\omega+E_d
\sim D\exp\biggl(-\frac{\pi\delta\omega}{2\Delta}\biggr).
\end{equation}
We see that the   perturbative expansion 
of the nonlinear optical polarization in terms  of the optical fields
{\em breaks down} even for weak pump 
intensities (i.e., small  $\Delta$). The  above condition
of its validity depends critically on the detuning of the pump
frequency from the Fermi level. 
For off-resonant pump, such that the effective impurity level
lies below the Fermi level, $|E_d|=|\varepsilon_d|-\omega_2\gg \Delta$,
the relation  (\ref{break}) can be written as
$\delta\omega+E_d\sim  T_K$ where 
\begin{equation}
\label{TK}
T_K=De^{\pi E_d/2\Delta}=
D\exp\biggl[-\frac{|\varepsilon_d|-\omega_2}{2g\mu^2{\cal E}_2^2}\biggr].
\end{equation}
This new energy scale can be associated with the Kondo 
temperature---an energy scale known to emerge from a spin-flip 
scattering of a FS electron by a magnetic
impurity\cite{hewson}. 
Remarkably, in our case, the Kondo 
temperature can be {\em tuned}  by varying the frequency and intensity
of the pump. In fact, the logarithmic divergence in
Eq. (\ref{alpha3}) is an indication of an {\em optically-induced}
Kondo effect. 

Let us now turn to the second term in Eq.\ (\ref{pol-kondo}). 
In fact, it represents the lowest order in the expansion of the 
linear polarization with impurity level shifted by 
$\delta\varepsilon=(N-1)\mu^2{\cal E}_2^2\sum_{\bf q}
\frac{n_q}{\varepsilon_q-E_d}$,
\begin{equation}
\label{pol-linear}
\tilde{P}^{(1)}=\mu^2 {\cal E}_1\sum_{\bf p}
\frac{(1-n_p)}{\varepsilon_p-\varepsilon_d+\delta\varepsilon-\omega_1}.
\end{equation}
% 
%where 
%$\delta\varepsilon=(N-1)\mu^2{\cal E}_2^2\sum_{\bf q}
%\frac{n_q}{\varepsilon_q-E_d}$. 
The origin of $\delta\varepsilon$ can
be understood by  observing that, for {\em monochromatic} 
pump, the coupling between the 
FS and the impurity can be described by a {\em time-independent}
Anderson Hamiltonian $H_A$ with effective impurity level
$E_d=\varepsilon_d+\omega_2$ and hybridization parameter $V=\mu{\cal E}_2$. 
By virtue of this analogy, $\delta\varepsilon$ itself is the
perturbative solution of the following equation for the self-energy 
part\cite{hewson}
\begin{eqnarray}
\label{eps0}
E_0=\Sigma(E_0)\equiv 
%&&
(N-1)\mu^2{\cal E}_2^2\sum_{\bf q}
\frac{n_q}{\varepsilon_q-E_d+E_0}
%\nonumber\\&&
\simeq (N-1)\frac{\Delta}{\pi}\ln \frac{E_d-E_0}{D},
\end{eqnarray}
which determines the renormalization of the effective impurity energy,
$E_d$, to $\tilde{E}_d=E_d-E_0$ \cite{hewson}.
Indeed, to the first order in the optical field, Eq. (\ref{eps0}) yields
$E_0=\delta\varepsilon$  after omitting
$E_0$ in the rhs.

The logarithmic divergence (\ref{alpha3}) indicates that near the
absorption threshold, a nonperturbative treatment is necessary.
Recall that, in the lowest order in $v_0$, the attractive interaction
$v_0$ between a localized hole and FS electrons also leads to a
logarithmically diverging correction even in the linear absorption:  
$\delta \tilde{P}^{(1)}\sim \tilde{P}^{(1)}gv_0
\ln[D/(\omega_1+\varepsilon_d)]$.
In the nonperturbative regime, 
$\delta \tilde{P}^{(1)}\sim \tilde{P}^{(1)}$,
this correction evolves into the Fermi edge
singularity\cite{mahan-book}. The question is how the Kondo 
correction (\ref{alpha3}) will evolve in the nonperturbative regime.
This question is addressed in the next section.

\section{general shape of absorption spectrum}

We start by discussing qualitatively the results and defer the details
to the end of the section.

It can be seen from the expression (\ref{TK}) for $T_K$ that there
is a well-defined critical pump intensity,
\begin{eqnarray}
\label{crit}
\Delta_c\equiv \pi g\mu^2{\cal E}_{2c}^2=
\frac{\pi}{2}\Bigl(|\varepsilon_d| -\omega_2\Bigr).
\end{eqnarray}
Note that $|\varepsilon_d| -\omega_2$ is the pump detuning from the
Fermi level.
The shape of the nonlinear absorption spectrum  depends sharply
on the ratio between $\Delta$ and $\Delta_c$. The strong pump case, 
$\Delta >\Delta_c$, is analogous to mixed-valence regime.
In this regime, the Kondo correction (\ref{alpha3}) develops
into a broad peak with width $\Delta$ and height $p_0$. This is
illustrated in Fig.\ 2(a).

Much more delicate is the case $\Delta\ll \Delta_c$,
which is analogous to the Kondo limit. 
The Kondo scale $T_K$ is then much smaller than $\Delta$,
which is the case for well-below-resonance pump excitation, 
$|\varepsilon_d|-\omega_2\gg \Delta$. The impurity density of states
in the Kondo limit is known\cite{hewson} 
to have two peaks well separated in energy by 
$|E_d|=|\varepsilon_d|-\omega_2 \gg\Delta$ ($E_d$ is the effective level
position). As a result, in the presence of the pump, the system sustains
excitations originating from the beats between these peaks.
These excitations can, in fact, assist the absorption of a probe
photon. The corresponding condition for the probe frequency reads
$|E_d|+\omega_1\simeq |\varepsilon_d|$, or
$\omega_1\simeq\omega_2$. Thus, in the Kondo limit, the
absorption spectrum exhibits a narrow peak below the linear
absorption onset. This is illustrated in Fig.\ 2(b). 

\begin{figure}
\begin{center}
\epsfxsize=3.0in
\epsffile{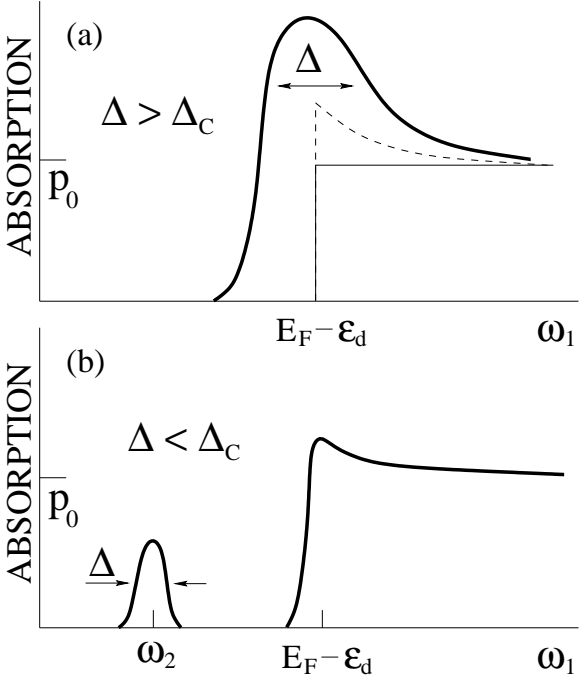}
\end{center}
%\vspace{0.5in}

\caption{Schematic plot of the Kondo-absorption spectra vs probe
frequency. (a) Mixed-valence regime: spectrum for strong pump
intensity (thick line) compared with the perturbative result 
(\ref{alpha3}) (dashed line) and the linear absorption spectrum (thin line).
(b) Kondo limit: the nonlinear absorption spectrum exhibits a narrow
peak below the linear absorption threshold.}
\end{figure}

To calculate the shape of the below-threshold absorption peak, we
adopt the large $N$ variational 
wave-function method by following the approach of \cite{gunn83}.  
For monochromatic optical fields,
the polarization (\ref{PP}) can be written as
\begin{equation}
\label{pol-green}
\tilde{P}=-\mu^2{\cal E}_1\Bigl[G^{<}(E_0-\delta\omega)
+G^{>}(E_0+\delta\omega)\Bigr],
\end{equation}
where
\begin{equation}
\label{green}
G^{<}(\varepsilon)=
\langle \Omega|T^{\dag}(\varepsilon -H_A)^{-1}T|\Omega\rangle,
\end{equation}
and $G^{>}(\varepsilon)$ is similar but with $T\longleftrightarrow T^{\dag}$.
In the leading order in $N^{-1}$, the ground state $|\Omega\rangle$
is given by 
\begin{equation}
\label{ground}
|\Omega\rangle= A\Bigl(|0\rangle 
+\sum_{\bf q}n_qa_q|{\bf q},1\rangle\Bigr),
\end{equation}
where 
$|{\bf q},1\rangle=
N^{-1/2}\sum_{\sigma}d_{\sigma}^{\dag}c_{{\bf q}\sigma}|0\rangle$
($|0\rangle$ stands for the full FS). The coefficients $A$ and $a_k$ are
found by minimizing $H_A$ in this basis,
\begin{equation}
\label{coeff}
A^2=1-n_d,
~~~
a_q=\frac{\sqrt{N}\mu{\cal E}_2}{E_d-\epsilon_q-E_0},
\end{equation}
where  
\begin{equation}
\label{occ}
n_d=\biggl(1+\frac{\pi \tilde{E}_d}{N\Delta}\biggr)^{-1}
\end{equation}
is the impurity occupation\cite{gunn83,hewson}
($N\Delta$ is finite in the large $N$ limit). Using that 
$\hat{T}|\Omega\rangle=\sqrt{N}\sum_{\bf q}|{\bf q},1\rangle$, the Green
function $G^{<}(\epsilon)$ takes the form
\begin{equation}
\label{green1}
G^{<}(\epsilon)=NA^2\sum_{{\bf pq}}n_pn_q
\Bigl\langle {\bf p},1 \Bigl|\frac{1}{\epsilon -H_A}
\Bigr|{\bf q},1\Bigr\rangle,
\end{equation}
where the matrix element in the rhs is obtained, to the leading
order in $N^{-1}$, by inverting $H_A$
in the above basis set,
\begin{eqnarray}
\label{matr}
%\langle {\bf p},1 |(\epsilon -H_A)^{-1}|{\bf q},1\rangle
\Bigl\langle {\bf p},1 \Bigl|\frac{1}{\epsilon -H_A}
\Bigr|{\bf q},1\Bigr\rangle
=\frac{\delta_{\bf pq}}{\epsilon+\epsilon_q-E_d}
+
%\frac{\Delta}{\pi}
\frac{1}{\epsilon-\Sigma(\epsilon)}
\frac{\frac{\Delta}{\pi}(N-1)}{(\epsilon+\epsilon_q-E_d)
(\epsilon+\epsilon_p-E_d)},
\end{eqnarray}
with $\Sigma(\epsilon)$ given by Eq. (\ref{eps0}). We then obtain
\begin{equation}
\label{green2}
G^{<}(E_0-\delta\omega)
=\frac{\pi}{\Delta}\Biggl[\Sigma(E_0-\delta\omega)
+
\frac{|\Sigma(E_0-\delta\omega)|^2}
{E_0-\delta\omega-\Sigma(E_0-\delta\omega)}
\Biggr].
\end{equation}
Since $\Sigma(E_0)=E_0$ [see Eq. (\ref{eps0})], the second 
term has a pole at $\delta\omega=0$ which gives rise to a resonance.
The $N^{-1}$ correction gives a finite resonance width $\Delta$.
The first term gives nonresonant contribution to absorption for
$\delta\omega<\tilde{E}_d$. In a similar way, it can be
shown that contribution from $G^{>}(E_0+\delta\omega)$
is suppressed by factor $N^{-1}$ and is nonresonant. 

Near the resonance, using that 
$[\partial\Sigma(E_0)/\partial E_0-1]^{-1}=n_d-1$ \cite{hewson}, 
we obtain

\begin{eqnarray}
\label{kondo-abs}
{\rm Im}\tilde{P}_K= 
p_0\,\frac{E_0^2(1-n_d)^2}
{\delta\omega^2+\Delta^2}.
% \sim 
% \biggl(\frac{\pi E_d T_K}{N\Delta}\biggr)^2
% \! \frac{p_0}{\delta\omega^2+\Delta^2}.
\end{eqnarray}
For $\Delta\ll \Delta_c$, corresponding to the  Kondo limit, 
we have $1-n_d\simeq \pi T_K/N\Delta$  and $E_0\simeq E_d$,
so that
\begin{eqnarray}
\label{kondo-abs-kondo}
{\rm Im}\tilde{P}_K
% \frac{p_0E_0^2(1-n_d)^2}
% {\delta\omega^2+\Delta^2}
\sim p_0\,\biggl(\frac{\pi E_d T_K}{N\Delta}\biggr)^2
\! \frac{1}{\delta\omega^2+\Delta^2}.
\end{eqnarray}
In this case, (\ref{kondo-abs-kondo}) describes 
the narrow below-threshold peak [see Fig.\ 2(b)]. 
The factor $(1-n_d)^2$ has the physical meaning of a
product of populations of electrons in the narrow peak of the impurity
spectral function (Kondo resonance) and ``holes'' in the wide peak
(centered at $\varepsilon_d$ below the Fermi level). 
In the opposite case, $\Delta\gtrsim \Delta_c$, 
corresponding to mixed-valence regime, we have 
$1-n_d\sim 1$ and $E_0 \sim N\Delta$. Then the polarization
\begin{eqnarray}
\label{kondo-abs-mixed}
{\rm Im}\tilde{P}_K= 
p_0\,\frac{N^2\Delta^2}
{\delta\omega^2+\Delta^2}
% \sim 
% \biggl(\frac{\pi E_d T_K}{N\Delta}\biggr)^2
% \! \frac{p_0}{\delta\omega^2+\Delta^2}.
\end{eqnarray}
describes the absorption peak in Fig.\ 2(a).

\section{coclusion}

Note that, although we considered here, for simplicity, the limit of 
singly occupied impurity level in the ground state, 
the Kondo-absorption can take place even if the impurity 
is {\em doubly} occupied. 
Indeed, after the probe excites an impurity electron,
the spin-flip scattering of FS electrons with the remaining 
impurity electron will lead to the Kondo resonance in
the final state of the transition. In this case, however, the Kondo
effect should show up in the fifth-order polarization.

A feasible system in which the proposed effect might be observed is,
e.g., GaAs/AlGaAs superlattice delta-doped with Si donors located in 
the barrier. The role of impurity in this
system is played by a shallow acceptor, e.g., Be. MBE growth 
technology allows one to vary the quantum well width and to place
acceptors right in the middle of each quantum well \cite{acceptor}.
In quantum wells, the valence band is only doubly degenerate with
respect to the total angular momentum J. Thus, such a system
emulates the large U limit considered here. The  dipole matrix
element for acceptor to conduction band transitions can be estimated
as $\mu\sim\mu_0 a$, where $\mu_0$ is the interband matrix element
and $a$ is the size of the acceptor wave function. 
For typical excitation intensities\cite{shah96}, the parameter
$\Delta$ ranges on the meV scale resulting in 
$T_K\sim \Delta$ for the pump detuning of several meV.

Finally, let us discuss the effect of a finite duration of the pump
pulse, $\tau$. Our result for $\chi^{(3)}$ remains unchanged if $\tau$
is longer than $\hbar/T_K$. 
If $\tau < \hbar/T_K$, then $\tau$ will serve as a cutoff 
of  the logarithmic divergence in (\ref{alpha3}), and the 
Kondo correction will depend on the parameters of the pump
${\cal E}_2$ and $\tau$ as follows: 
$ {\rm Im}\tilde{P}_K^{(3)}\propto {\cal E}_2^2\ln(D\tau/\hbar)$.
In the non-perturbative regime, our basic assumption was that, for
monochromatic pump, the system maps onto the {\em ground} state of the
Anderson Hamiltonian. Our results apply if the pump is
turned on slowly on a time scale longer than
$\hbar/T_K$. For shorter 
pulse duration, the build up of the optically-induced Kondo effect
will be determined by the dynamics of FS excitations.
% The role of interactions between FS and impurity electrons
% in the presence of hybridization was addressed in 
% \cite{perakis93}. 
% An avenue for future studies would be the
% interplay between the Kondo-absorption and the Fermi edge singularity.
%
%Note finally that the effect of irradiation on the Kondo transport in
%quantum dots was investigated in \cite{glazman99,dots}.

This work was supported by ONR/DARPA (TVS and IEP),
and by NSF grant INT-9815194 and Petroleum Research Fund grant
ACS-PRF \#34302-AC6 (MER).


\begin{references}

\bibitem{chemla99} See, e.g.,  
D. S. Chemla, in 
{\em Nonlinear Optics in Semiconductors}, edited
by R.\ K.\ Willardson and A.\ C.\ Beers\ (Academic\ Press,\ 1999).

\bibitem{shah96} See, e.g.,
J.\ Shah, {\em Ultrafast Spectroscopy of Semiconductors and
Semiconductor nanostructures} (Springer, 1996).

\bibitem{perakis00} See, e.g.,
I.\ E.\ Perakis and T.\ V.\ Shahbazyan, 
{\em Surf.\ Sci.\ Reports}\ {\bf 40}, 1 (2000).

\bibitem{bre95} 
I.\ Brener, W. H. Knox, and W. Schaefer,
{\em Phys. Rev. B} {\bf 51}, 2005 (1995);
I.\ E.\ Perakis,
I. Brener, W. H. Knox, and D. S. Chemla,
{\em J.\ Opt.\ Soc.\ Am.\ B}\ {\bf 13},\ 1313\ (1996).

\bibitem{portengen}T. Portengen, Th. Ostreich, and L. J. Sham, 
{\em Phys. Rev. Lett.} {\bf 76}, 3384 (1996); 
{\em Phys. Rev. B} {\bf 54}, 17452 (1996).

\bibitem{awschalom}
J. M. Kikkawa and D. D. Awschalom, 
{\em Phys.\ Rev.\ Lett.}\ {\bf 74},\ 80\ 4313 (1998).


\bibitem{hall}
N. A. Fromer, C. Schüller, D. S. Chemla,
T. V. Shahbazyan, I. E. Perakis,
K. Maranowski, and A. C. Gossard 
{\em Phys.\ Rev.\ Lett.}\ {\bf 83},\ 4646\ (1999). 

\bibitem{dodge99}
J. S. Dodge, A. B. Schumacher, J.-Y. Bigot, D. S. Chemla,
N. Ingle, and M. R. Beasley, 
{\em Phys.\ Rev.\ Lett.}\ {\bf 83},\ 4650\ (1999).

\bibitem{FES}
N.\ Primozich,
T. V. Shahbazyan, I. E. Perakis, and D. S. Chemla, 
{\em Phys.\ Rev.\ B}\ {\bf 61},\ 2041\ (2000);
T.\ V.\ Shahbazyan,
N. Primozich, I. E. Perakis, and D. S. Chemla, 
{\em Phys.\ Rev.\ Lett.}\ {\bf 84},\ 2006\ (2000).

\bibitem{mahan-book} 
G.\ D.\ Mahan, 
{\em Many-Particle Physics}\ (Plenum,\ 1990).
%\ pp.\ 732-764.


\bibitem{mukamel-book}S.\ Mukamel, 
{\em Principles of Nonlinear Optical Spectroscopy}
(Oxford University Press, 1995).

\bibitem{shahbazyan00}
T. V. Shahbazyan, I. E. Perakis, and M. E. Raikh,
{\em Phys.\ Rev.\ Lett.}\ {\bf 84},\ 5896\ (2000).



\bibitem{hewson}A.\ C.\ Hewson, 
{\em The Kondo Problem to Heavy Fermions}
(Cambridge University Press, 1993).


\bibitem{gunn83}O.\ Gunnarson and K.\ Sch\"{o}nhammer,
{\em Phys.\ Rev.\ B}\ {\bf 28},\ 4315\ (1983).

\bibitem{acceptor}
P. O. Holtz, Q. X. Zhao, B. Monemar, M. Sundaram, J. L. Merz, and
A. C. Gossard,
{\em Phys.\ Rev.\ B}\ {\bf 47},\ 15 675\ (1993);
I. V. Kukushkin,
K. von Klitzing, K. Ploog, and V. B. Timofeev,
{\em Phys.\ Rev.\ B}\ {\bf 40},\ 7788\ (1989).

% \bibitem{perakis93}I. E. Perakis,
% C. M. Varma, and A. E. Ruckenstein, 
% {\em Phys.\ Rev.\ Lett.}\ {\bf 70},\ 3467\ (1993); 
% I.\ E. Perakis and C.\ M. Varma, 
% {\em Phys.\ Rev.\ B}\ {\bf 49},\ 9041\ (1994).


%\bibitem{glazman99}A. Kaminski,
%%\ {\em et\ al.},
%Yu. Nazarov, and L. I. Glazman,
%Phys.\ Rev.\ Lett. {\bf 83},\ 384\ (1999).

%\bibitem{dots}M.\ H.\ Hettler and H.\ Sch\"{o}ller, 
%Phys.\ Rev.\ Lett.\ {\bf 74},\ 4907\ (1995); 
%T.\ K.\ Ng, {\em ibid.}\ {\bf 76},\ 487\ (1996); 
%R.\ Lopes\ {\em et\ al.},\ {\em ibid.}\ {\bf 81},\ 4688\ (1998);
%Y.\ Goldin and Y.\ Avishai,  {\em ibid.}\ {\bf 81},\ 5394\ (1998);
%P. Nordlander\ {\em et\ al.},\ {\em ibid.}\ {\bf 83},\ 808\ (1999).

\end{references}
\end{document}